\documentclass{aa}

\usepackage{graphics}

\begin{document}


   \title{Collisions between carbonaceous grains in the interstellar medium}
\author{R. Papoular
\inst{1}
          }

   \offprints{R. Papoular}

   \institute{Service d'Astrophysique and Service de Chimie Moleculaire, CEA Saclay, 91191 Gif-s-Yvette, France\\
              e-mail: papoular@wanadoo.fr
             }

   \date{ }
   \authorrunning {R. Papoular}
   \titlerunning {Grain collisions}
   \maketitle
\begin{abstract}

Semi-empirical molecular dynamics is used to simulate hydrocarbon grain sputtering and collisions which are extremely difficult to study experimentally. This microscopic and dynamic approach is particularly suited to high velocity impacts, where target destruction occurs far from equilibrium. A wide variety of processes are encountered, depending on grain size and velocity: vaporization, fragmentation, atomic implantation, sticking, elastic recoil, atomic chemisorption, H abstraction and H$_{2}$ formation, etc. The impact-velocity threshold is about 10 km/s but complete grain destruction requires much higher velocities and nearly equal grain sizes. The main outcome of strong collisions is vaporization, i.e. formation of small molecules rather than solid fragments. Most of the impact energy is carried away by these molecules in the form of kinetic energy, with the consequence that the destruction efficiency of collisions is considerably reduced.

 \keywords{ISM: dust, ISM:kinematics and dynamics, molecular processes}
 %
  \end{abstract}

%
\section{Introduction}

Strong collisions between IS (InterStellar) dust grains are commonly expected to alter significantly, if not outright determine, the grain size distribution, mainly by destroying big grains thereby creating small ones, but perhaps also by creating large grains through coagulation of smaller ones (see Wickramasinghe \cite{wic67}, Bierman and Harwitt  \cite{bier80},  Spitzer  \cite{spi78}). If grain destruction is predominant in the ISM (IS Medium), then, grains larger than 1000 \AA{\ }are expected to be almost depleted; since they are observed to be quite abundant, the far-reaching conclusion may also be drawn, that grains must also be able to grow even in such a tenuous environment (see Boulanger and Cox  \cite{bou98}; Miville-Deschenes et al. \cite{miv02}). Also, if collisions induce vaporization, new molecules can be created, thus contributing to the rich population of observed molecules in the ISM. Finally, even if shattering does not occur upon collision, intense heating may cause structural changes in the grains. These expectations are significant enough to have triggered a lasting interest in grain collisions for at least half a century (see Salpeter  \cite{sal74}; Barlow and Silk \cite{bar77}; Cowie \cite{cow78}; Spitzer  \cite{spi78} and Draine and Salpeter  \cite{dra79}) and produced a large body of literature. Since grains of relevant size are very difficult to produce, handle and accelerate in the laboratory, it is not surprising that this literature is essentially theoretical in nature, although, of course, it does rely upon whatever experimental results are available.

Earlier treatments considered vaporization to be the dominant result of grain collisions, and to be complete beyond the relative velocity at which the projectile kinetic energy was equal to the total dissociation energy. Assuming the collision is totally inelastic and all the kinetic energy is available for dissociation, this velocity is independent of grain size and estimated at 3 km/s for icy dielectrics up to 8 km/s for iron particles (Spitzer  \cite{spi78}). More recent papers on the subject have relied upon condensed-state shock theory in order to predict effects of grain collisions (see Borkowski and Dwek  \cite{bor95}, Tielens et al. \cite{tie94}, Seab \cite{seab90}). This approach enhances shattering and fragmentation over vaporization, with a consequent lowering of grain destruction threshold down to relative grain velocities of about 1 km/s, and complete target disruption at 100 km/s. Admittedly, however, such treatments apply best to macroscopic bodies and are not easily extended to the grains of interest here. Moreover, little data is available on shocks in materials similar to those that are thought to be the constituents of IS grains.

The theoretical and experimental obstacles encountered in the field of small grain collisions, even outside the realm of astronomy, and the development of material and chemical computation software have prompted a flurry of numerical simulations which shed a striking new light on the subject (see Cleveland and Landeman  \cite{cle92}; Schulte \cite{schu95}; Aoki et al. \cite{aok97}). Of course,because of limitations on available computer memory, this approach is best suited to relatively small systems, but these are precisely those we are interested in. Moreover, the composition and structure of the interacting systems may be tailored to suit the problem at hand. Also, various relevant microscopic phenomena can be followed quantitatively in time and space with  exquisite detail: atomic adsorption, abstraction and implantation, molecule formation, grain cratering, shattering or vaporization, etc (see Davenas \cite{dav93}). Again, by contrast with shock theory, which subsumes thermal quasi-equilibrium,  the time scales of computations can be short enough (10$^{-16}$s) to accomodate the out-of-equilibrium phenomena that are expected to occur upon high-velocity impacts of small systems.

These potentialities are taken advantage of to revisit grain collisions in the present paper. The scope here is limited to carbonaceous material similar to the presently envisioned constituent of IS grains: partly aromatic (6- and 5-C rings), partly aliphatic (C-H chains) and more or less hydrogenated. Their sizes range from 1 to 60 atoms and their velocities from 1 to 300 km/s. The penetration range of \emph{single} C and H atoms in this material is first determined as a function of impact velocity. This already gives some insight into the mechanism and extent of material destruction upon collision. More quantitatively, a cross-section for bond dissociation, and the corresponding threshold, are determined as a function of impact velocity.

Impacts of \emph{clusters} of the same model material are then analyzed, which provides the average dissociation energy of the material, as well as rudimentary statistics of product sizes. These results are finally used to estimate the ranges of grain size and velocity in which strong size modifications are to be expected. Generally speaking, these estimates are less pessimistic than is usually the case for the survival of C-rich grains in the ISM. However, no discussion is offered here of the detailed effects of collisions on grain size distributions.

\section{Designing the collision experiment}
A description of the software package used for simulation can be found in Papoular (\cite{pap02}). The particular code used here is AM1 (Austin Model 1), as proposed by Dewar et al. (\cite{dew85}). This semi-empirical method combines a rigorous quantum-mechanical formalism with empirical parameters obtained from comparison with experimental results. It computes approximate solutions of Schroedinger's equation, and uses experimental data only when the Q.M. calculations are too difficult or too lengthy. This makes it more accurate than poor ab initio methods, and faster than any of the latter.

First, a molecule of the desired composition, structural type and size is built on the computer screen by picking each atom in turn from a table of elements. The code is then used to optimize the molecular geometry, i.e. determine the nearest architecture which minimizes the total potential energy of the system. Here, the initial spatial distribution of atoms was chosen so that, upon optimization, the structure included short chains as well as 5- and 6-membered rings of carbon atoms. Hydrogen atoms were then added randomly to simulate partial hydrogenation. At the end of the process, the distances between atoms are indeed those expected for hydrocarbons. The process is repeated to form the second reactant cluster of the collision. The cluster sizes adopted here are a compromise between computational time and relevance to real IS dust.

Then, the reactants are set at an initial distance such that interaction between them is negligible: more than 3 \AA. Finally, one of them, the projectile, is given, as a whole, the desired velocity towards the other, the target; the latter is given velocity 0 and the calculation is started. In the example of fig. 1, the target is a 60-atom cluster and the projectile, on its right side, is comprised of a single C atom.

The molecular dynamics relies on the Born-Oppenheimer approximation to determine the movements of atoms under nuclear and electronic forces due to their environment; the calculation is made by steps of 10$^{-4}$ ps and 10$^{-3}$ ps respectively for high and low relative velocies. At every step, all the system parameters are memorized as snapshots so that, after completion of the run, a movie of the collision can be viewed on the screen, and the trajectory of any atom followed all along. This makes for a better understanding of the details of destruction mechanisms and outcomes. Note that the dynamics of bond dissociations can only be simulated by using Unrestricted Hartree-Fock (UHF) wave functions in the Q.M. part of the calculation (see Szabo and Ostlund \cite{sza89}).

\begin{figure}
\resizebox{\hsize}{!}{\includegraphics{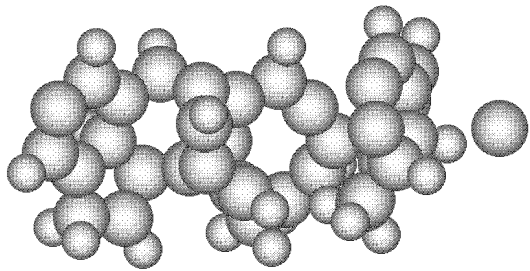}}
\caption[]{Set up of a collision simulation. The lone C atom (right) is the projectile, with an initial velocity directed along the primary axis of the target.
Target (left) made of amorphous hydrocarbon material: 36 C atoms (larger spheres) linked in chains, 5- and 6-membered rings, and 24 H atoms occupying some of the available carbon valences. Length $\sim$12 \AA, height and depth $\sim$8 \AA.}
\end{figure}

\section{Atomic carbon projectile}
A projectile made of a single C atom, as in fig.1, is a good start since it is easier to follow its trajectory and energy after it hits the target. Also, its behaviour already encapsulates the main features of grain collisions for, at the high velocities of interest here, the translational energy of a projectile atom is much larger than its binding energy, so that, even if it is part of a cluster, it behaves nearly independantly of its neighbours.

At the beginning of each numerical experiment in this series, the target will be set at rest and the projectile, a few angstroms away, will be given a velocity $V_{\rm i}$ towards the target. Since gaseous flows in the ISM usually enforce relative velocities between grains, $V_{\rm i}$ will be taken as the main control parameter rather than the kinetic energy. $V_{\rm i}$ will be expressed in units of \AA/ps (0.1 km/s). It is found that, for $V_{\rm i}$ $\leq$100 \AA/ps, the C atom most often simply sticks onto the nearest ``acceptor" site in the target. This is a result of the strong reactivity of atomic carbon, which has four dangling (free) bonds, as well as of the reactivity of the target, whose carbon bonds are not all saturated with hydrogen. In such a sticking event, the target structure is hardly modified, except, perhaps, for the displacement of one of its hydrogen atoms. Incidentally, this sticking process is also an elementary step in the process of grain growth. Not being interested, at  this time, in events of this type, we now focus on the velocity range beyond 100 \AA/ps (10 km/s). Note that at this velocity a C atom carries a kinetic energy of 6.24 eV, somewhat higher than the bond energy in hydrocarbons. By way of comparison, Salpeter (\cite{sal74}) set the threshold velocity for \emph{sputtering} by heavy atoms at $\sim$30 km/s.

\begin{figure}
\resizebox{\hsize}{!}{\includegraphics{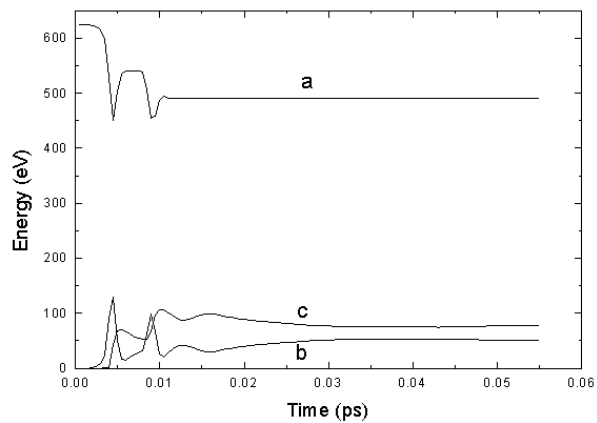}}
\caption[]{Energetics of a collision experiment (configuration of fig. 1 with initial projectile velocity=1000 \AA/ps). a) K.E. of the projectile; b) variation of the target P.E.; c) (thick line) K.E. of the target. The two pairs of opposite narrow peaks correspond to close encounters between the projectile and one target C atom. Here, following one such collision, the projectile left the target prematurely at 0.01 ps with velocity 887 \AA/ps. Exchanges between target K.E. and P.E. are still visible afterwards.}
\end{figure}

Figure 2 applies to the case of $V_{0}$=1000 \AA/ps (100 km/s). Plotted as a function of time are the kinetic energy of the projectile (a), that of the target (c), and the \emph{variation} of the target potential energy (b), throughout the ``reaction". The interpretation of these curves is made easier if, after the computer run, the accelerated motion of the computation is replayed on the screen and snapshots of the system are taken at characteristic times along the x-axis of fig. 2. It is found that the projectile hits the target at t=0.0025 ps and leaves it at t=0.01 ps, after describing an irregular trajectory about 7 \AA{\ }in length. Large path deviations occur at 0.0039 and 0.0085 ps, when the projectile experiences a close encounter with a C atom of the target. In such events, the projectile is strongly decelarated but soon recovers most of its velocity because, in the present range of velocities, the dominant interaction with the target atoms is a ``quasi-elastic" recoil of nuclei (see Sigmund \cite{sig69}). Hence the dips in projectile K.E. (kinetic energy). Correspondingly, the target momentarily gains P.E. (potential energy): peaks of curves (a) and (b) are of opposite signs. However, all the while the projectile is inside the target, the general trend is a reduction of its velocity as it loses energy to bordering target atoms in the form of momentum. Hence the initial rise in target K.E. (curve c). Part of this gain is readily converted to target P.E, leading to a concomitant rise in curve (b). For impact parameters such that the energy transferred from projectile to target atom is larger than the bond energy, part of this energy goes into breaking a bond, and the rest into translational and rotational K.E., as well as internal energy (vibrational K.E. and P.E. and, possibly, deformation or restructuring energy) of the fragments produced. Hence the small bumps of opposite signs in curves (b) and (c), accompanying the large recoil peaks. 

In the end, only about 30 \% of the projectile energy was absorbed by the target. This is only due to the  accidental early exit following a close encounter with a heavy target atom, and it will be shown below that the stopping range for this initial velocity is much larger than the present target thickness. In the case at hand, internal vibrations are only weakly excited in the fragments, as evidenced by the smoothness of the later parts of curves (b) and (c). However the final gain in potential energy is 50 eV. Most of this was used to dissociate the 6 observed fragments. The rest went into structural changes in the fragments as compared to the original material.

Note that the process of fragmentation and deformation occured in less than 0.04 ps, i.e. much less than the thermalization time (numerically determined for this type of material to be longer than 1 ps; see Papoular \cite{pap02}), and even shorter than the transit time of a shock wave through the target. It is clear, therefore, that thermodynamics cannot apply here; indeed, the sputtering phenomenon is usually treated at the microscopic level  (see Sigmund \cite{sig69}
 and Schiott \cite{sch68}).
 
Experiments at lower initial projectile velocities show that, as $V_{i}$ decreases, the number of fragments produced and the residual projectile velocity when it leaves the target both decrease. When $V_{i}\leq$300 \AA/ps, the projectile stops within the target, at a depth which decreases with the initial velocity. Also, its trajectory inside the target becomes more and more wrinkled and ends up as a random walk until it finally settles in some available ``acceptor" site (dangling bond or defect site). This, then, is a case of \emph{atomic implantation}.

In order to quantify this behaviour, we proceed by analogy with the usual method, which assumes that the loss, $dE$, of projectile K.E. is proportional to the density of target nuclei and to the linear element, $dx$, of its trajectory: $-dE/E=n \sigma dx$, where $\sigma$ is a cross-section describing the interaction. Here, we replace the denominator E by $\epsilon$, a typical bond energy: this ensures that the dimension of $\sigma$ is still that of an area and, at the same time, clarifies the physics, since the left hand side now represents the energy loss normalized by a characteristic energy of the system. Then, if $m$ is the projectile mass, and $V$ its velocity
\begin{equation}
-V dV=(n\epsilon/m) \sigma dx.
\end{equation}

The target mass does not appear here because it is assumed to be much larger than that of the projectile. The results of the numerical simulations, as well as the physics of these experiments, are incompatible with a constant cross-section. But agreement is good enough if it is assumed that : $\sigma=a+bV+cV^{2}$. Clearly, the dimensions of $a$, $b$ and $c$ are, respectively, those of an area, an area per velocity and an area per velocity square. If not stated otherwise, the units of length, area and velocity will be, respectively, \AA, \AA$^{2}$ and \AA/ps. Then, the solution of eq. 1 is of the form

\begin{equation}
x_{\rm f}-x_{\rm i}=\frac{n\epsilon\gamma}{m}[\alpha\ln(\frac{ V_{\rm i}-\alpha}{ V_{\rm f}-\alpha})+\beta\ln(\frac{ V_{\rm f}-\beta}{ V_{\rm i}-\beta})]
\end{equation}

where i and f designate initial and final values, and $\alpha$, $\beta$ and $\gamma$ (in the same units as the velocity $V$) are functions of $a, b$ and $c$. We seek values of $\alpha$, $\beta$ and $\gamma$ so that eq. 2 is best satisfied by the triads, $V_{\rm i}$, $V_{\rm f}$ and $\Delta$$x=x_{\rm f}-x_{\rm i}$, corresponding to this series of 7 numerical  experiments at different $V_{\rm i}$'s: 150, 200, 300, 400, 500, 750 and 1000 \AA/ps. We find: $\alpha$=1250 , $\beta$=-130  and $\gamma=5.6 10^{-2}$  . Deriving $a$,$b$ and $c$ from these values of $\alpha$, $\beta$ and $\gamma$, and substituting in the quadratic expression for $\sigma$, it is finally found that  eq. 2 is satisfied to a standard deviation of 2 \% if 

\begin{equation}
\sigma=6.6+5.1 10^{-2}V-4.1 10^{-5}V^{2}
\end{equation}

where $\sigma$ is in \AA$^{2}$, $V$ in \AA/ps, and we adopted $n=10^{23}$ cm$^{-3}$, $m$=12 atomic units and $\epsilon$=5 eV.

\begin{figure}
\resizebox{\hsize}{!}{\includegraphics{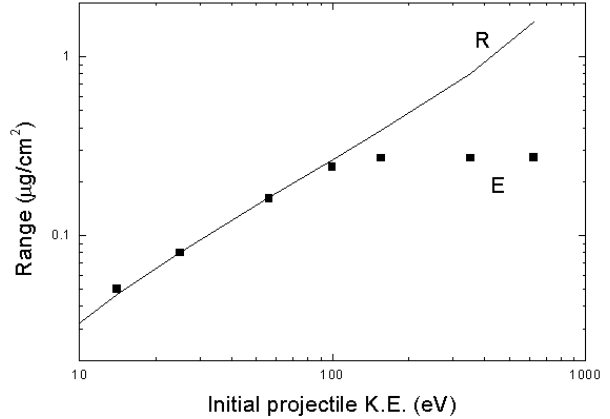}}
\caption[]{Line R: range of a carbon atom in the model material of fig.1, as a function of its initial K.E., according to eq. 2 and assuming a material density of 2 g/cm$^{3}$; R($\mu$g/cm$^{2}$)=0.02 $\Delta$$x$ (\AA); squares (E): range measured in individual simulations at discrete initial velocities.}
\end{figure}

Note that $\sigma$ is of the order of a few atomic radii squared, meaning that, in this range of velocities, the projectile hardly interacts with target atoms off its trajectory by more than 4 \AA. Also, from eq. 2, an expression can be derived for the range of the projectile as a function of its initial velocity. Indeed, when the projectile's trajectory ends within the target, its final velocity can be defined by the condition that it become random. In the present experiments, this is found empirically to be of order 50 \AA/ps, which will be adopted for $V_{\rm f}$. The range, $\Delta$$x$ is obtained by inserting this value in eq. 2.  While $\Delta$$x$ is a length, experimental studies with different materials show that the energy loss is less intimately connected with the length of the particle trajectory in the target than with the mass of target material swept by the projectile along its path, which we will denote by R and is equal to $\rho.\Delta$$x$, where $\rho$ is the density of the target material. If $\rho$=2 g/cm$^{3}$ and R is in $\mu$g/cm$^{2}$, then R=0.02 $\Delta$$x$. This awkward dimension for R is universally used by experimentalists because it ensures that the corresponding numerical values can be handled more easily, as illustrated by fig. 3, which represents the present results in these units, and can therefore be compared directly with published experiments. The nearly linear functional form as well as the absolute values of this curve are in line with previous sputtering studies (see Schiott \cite{sch68} for instance) and numerical simulations (see Cleveland and Landeman  \cite{cle92}; Aoki et al.  \cite{aok97}). The distances travelled by the projectile inside the target in the 7 simulations are also plotted as a scatter graph (E) in fig. 3. Beyond 400 \AA/ps (K.E.$\sim$100 eV), the range according to eq. 2 is longer than the long dimension of the target (12 \AA) but, of course, the ordinates of  ``experimental" points cannot reach much higher than this value, so they fall below the curve.

Because of the discrete nature of the target material structure, $\Delta$$x$ and $V_{\rm f}$ depend slightly (and randomly) on the exact direction of the initial projectile velocity, so that, even below 100 eV, there is some scattering of points E about curve R. Other illustrations of this scattering are provided by fig. 8 to 10. Such scattering is limited if the grain structure is not too anisotropic and if care is taken to avoid head-on atomic collisions and subsequent premature exit of the projectile from the finite target (cf fig. 2).

\section{Atomic hydrogen projectiles}

Several simulations were made with initial velocities from 25 to 3000 \AA/ps (2.5 to 300 km/s) and various impact angles on targets of the same structure as above. In none of these cases did fragmentation occur.  Below 30 \AA/ps, the projectile simply \emph{recoiled} from the target. Between 50 and 100 \AA/ps, the privileged outcome was \emph{abstraction} of a target H to form a hydrogen molecule; between 100 and 500 \AA/ps, \emph{chemisorption} on the target was preferred. Finally, at 1000 \AA/ps and higher velocities, the projectile traversed part or all of the target, leaving only a small fraction of its energy in the form of target vibrations. 

The maximum projectile energy in these simulations was about 500 eV. At this energy, the sputtering yield of H$^{+}$ on graphite was measured to reach a maximum of only 0.1 (Vietzke and Philipps \cite{vie89}). Thus, H atoms are not expected to contribute significantly to the destruction of grains; the physical reason for this is that their mass is so low (Salpeter \cite{sal74}).

\section{Cluster projectiles}

Armed with an understanding of the behaviour of single-atom projectiles, we can now turn to projectile sizes of the same order as that of the target. Figure 4 shows a series of snapshots of target and projectile taken during their collision. Each is made of 24 C and 12 H atoms, forming a structure similar to that of fig. 1; the initial velocity of the projectile (right of fig. 4) is 1000 \AA/ps towards the target (left), which is 11 \AA{\ }away . At t=0.01 ps (fig. 5), the projectile is entirely inside the target, with no dramatic change observed in either structure. At this stage, momentum has been exchanged but relative atomic distances within each grain are still nearly unchanged. At t=0.02 \AA {\ }(fig. 6), projectile and target are still recognizable but they have traded sides and each has ``expanded": dissociation has begun, but only after the projectile has made its way through, and out of, the target. This was to be expected from fig. 3, which indicates, for a carbon projectile of velocity 1000 \AA/ps, a range of about 50 \AA, much longer than the present target thickness (3.5 \AA). As to the target, note that its center of mass has barely moved in the laboratory frame up to this time. Much later, at t=0.1 ps (fig. 7), projectile and target are shattered in 49 distinct fragments gathered in 2 clusters of nearly equal sizes, and whose centers of mass are separated by about  80 \AA.
\begin{figure}
\resizebox{\hsize}{!}{\includegraphics{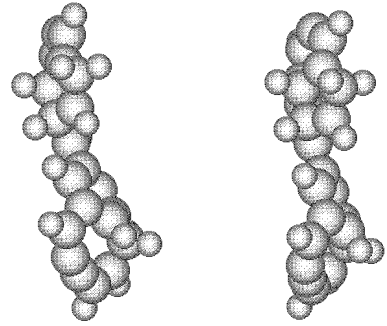}}
\caption[]{Snapshot of a collision between two clusters of equal sizes (40 atoms) at t=0: projectile (right) with initial velocity 100 km/s towards the target (left); distance between centers of mass: 11 \AA.}
\end{figure}
\begin{figure}
\resizebox{\hsize}{!}{\includegraphics{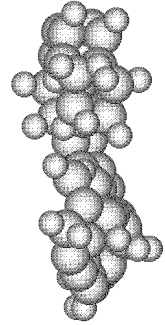}}
\caption[]{Snapshot of a collision between two clusters of equal sizes (40 atoms) at t=0.01 ps: projectile and target overlap.}
\end{figure}
\begin{figure}
\resizebox{\hsize}{!}{\includegraphics{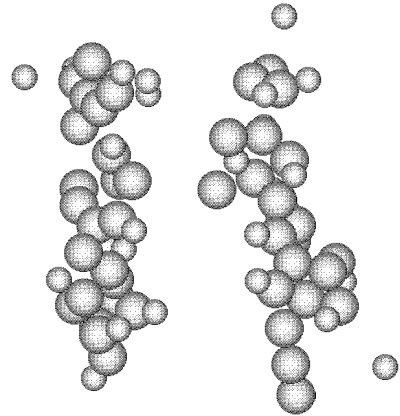}}
\caption[]{Snapshot of a collision between two clusters of equal sizes (40 atoms) at t=0.02 ps: projectile now on the left after traversing the target, distance: 9 \AA, dissociation just begun.}
\end{figure}
\begin{figure}
\resizebox{\hsize}{!}{\includegraphics{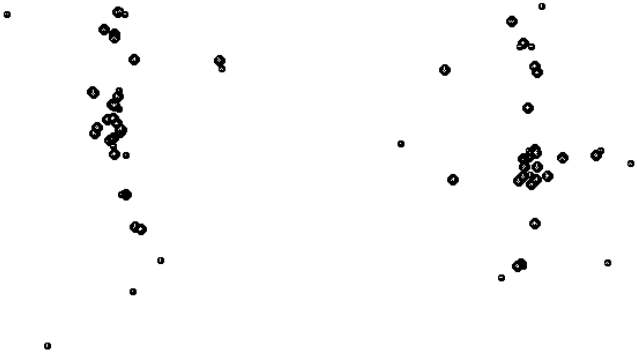}}
\caption[]{Snapshot of a collision between two clusters of equal sizes (40 atoms) at t=0.1 ps: nearly complete vaporization, 49 fragments, mean distance between clusters $\sim80$ \AA. Note that the scale is much larger than for fig. 4 to 6.}
\end{figure}
Before we discuss these fragments in detail, consider first the energetics of the collision. Throughout the process, and similar to fig. 2, total kinetic and potential energy vary in opposite directions and total energy is conserved to the accuracy of the calculation. At t=0, K.E.=15816 eV, entirely associated with the projectile. At t=0.01 ps, the clusters are spatially overlapping and the total gain in P.E. peaks at about 1400 eV, with a corresponding negative peak in K.E. (deceleration of projectile by the potential barrier of the target atoms). The full width of these peaks is 0.007 ps, which is the time spent by the projectile inside the target for this initial velocity. During this time, a fraction of the translational K.E. of the projectile is converted into internal K.E. in both clusters. This energy, in turn, is expended in bond dissociations and transformed in P.E. and K.E. of the fragments. At the end of the reaction, the two groups of fragments nearly equally share the total gain in P.E., which is $\sim$262 eV. For 49 products (fig. 7), this amounts to an average of 5.3 eV per dissociation. On the other hand, the final K.E.s of the remains of target and projectile are very different: the latter have retained $\sim$88$\%$ of the initial K.E. while the former have only gained 61 eV; and the velocity of the latter's center of mass is still a full 940 \AA/ps (in the initial direction), while the velocity of the former's center of mass is only 60 \AA/ps (nearly in the same direction).

Similar numerical simulations were performed for smaller and slower projectiles. Sizes of 40 (24 C and 16 H), 20 (12 C and 8 H), 10 (6 C and 4 H), 4 (2 C and 2 H) atoms as well as a single C atom (Sect. 3) were used. The corresponding set of results is represented in fig. 8, where the abscissa is the initial projectile velocity and the ordinate is the ratio of the number of atoms in fragments dissocited from target and projectile, to the total number of atoms in both clusters: this is taken as measure of the dissociation efficiency of a collision. The following can be noted:

\begin{figure}
\resizebox{\hsize}{!}{\includegraphics{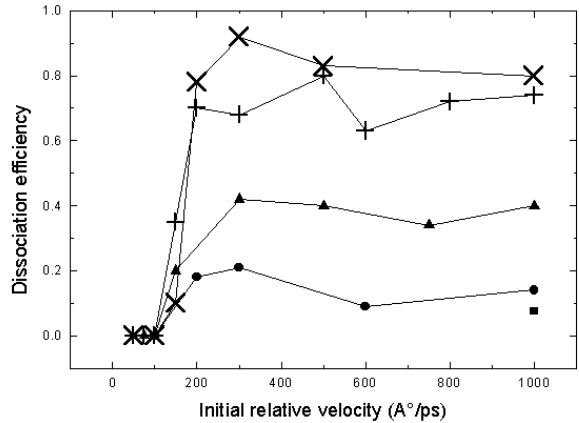}}
\caption[]{Dissociation efficiency of collisions: ratio of number of atoms in fragments dissociated from target and projectile, to total number of atoms in both clusters; as a function of initial relative velocity. Crosses, plusses, up triangles, circles and a square are used, respectively, for projectiles of 40 (24 C and 16 H), 20 (12 C and 8 H), 10 (6 C and 4 H), 4 (2 C and 2 H) atoms as well as a single C atom; the target is a 40-atom cluster ( same as in fig.4) for all simulations in this figure. Targets and projectiles (except single C atom) have same thickness, 3.5 \AA, parallel to the initial velocity and to the line joining centers of mass. The size of projectiles varies in the plane perpendicular to this direction. To help the eye, lines are drawn between discrete simulation results. 1000 \AA/ps=100km/s.}
\end{figure}

\begin{itemize}
\item
For all particle sizes, dissociation begins at $V_{\rm i}$$\sim$100 \AA/ps (10 km/s), when each C atom of the projectiles carries $\sim$6 eV of energy, just enough to break a carbon bond.

\item
For all projectile sizes, a maximum number of dissociations is essentially achieved  at $V_{\rm i}$$\sim$200 \AA/ps, when according to fig. 3, the range of a C atom into the target is just about equal to  the thickness of target or projectile in the present series of experiments (3.5 \AA). Beyond this, unused energy is carried away by the remains of the projectile leaving the target.

\item
The maximum number of dissociations that is achieved increases with the transverse size (hight) of the projectile relative to that of the target, but not much beyond a size ratio of $\sim$0.5. In this series of simulations, this \emph{ratio} corresponds to a \emph{difference} of transverse sizes of only 10 \AA, or about the cross-section diameter of a carbon atom suggested by the results of Sect. 3. Thus, it is likely that, for grain sizes much larger than those considered here, maximum dissociation efficiency would be reached only for nearly equal transverse sizes of projectile and target.

\item
Comparison of simulation results suggests that, as far as dissociation efficiency is concerned, the effect of a projectile can be considered as the sum of the effects of its individual carbon atom constituents. For instance, for given target size and initial projectile velocity, the final gain in total P.E. (which is proportional to the number of resulting dissociations) is nealy proportional to the number of C atoms in the projectile.
\end{itemize}
Drawing upon the same set of simulations, fig. 9 plots the number of products (= number of broken bonds) as a function of the final gain in total P.E. The average ratio of these quantities is found to be 5.6 eV per bond. On the other hand, the average bond energy of the model material may be estimated as the ratio of total initial P.E. (9227 kcal or 401 eV) to total number of atoms (80), giving 5 eV per atom, which is lower than the average dissociation energy just derived. The difference of 0.6 eV per dissociation indicates that the fragments also carry some vibrational energy, which amounts here to an average temperature of at most 4660 K divided by the number of atoms in a fragment. All this vibrational energy is but a tiny fraction of the initial projectile K.E.; it increases as the latter decreases towards the threshold ($V_{\rm i}$$\sim$100 \AA/ps) but always remains small.

\begin{figure}
\resizebox{\hsize}{!}{\includegraphics{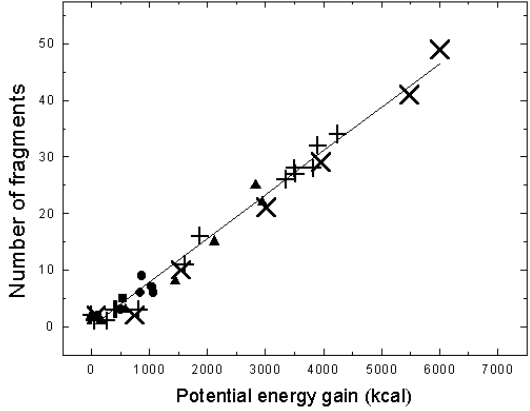}}
\caption[]{The average final gain in total P.E. per dissociation (or broken bond) is the inverse slope of the straight line fit: 5.6 eV per bond. 1eV=23 kcal.}
\end{figure}

\begin{figure}
\resizebox{\hsize}{!}{\includegraphics{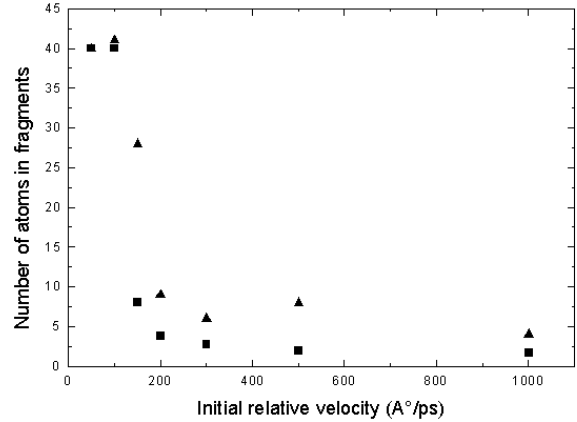}}
\caption[]{Maximum (triangles) and average (squares) number of atoms in fragments resulting from a collision between clusters of equal sizes (40 atoms), as a function of the initial relative velocity. The fragments are essentially small molecules as soon as the range corresponding to the relative velocity exceeds the size of the clusters. 1000 \AA/ps=100km/s.}
\end{figure}

Other interesting results of these experiments involve the average and maximum sizes of the products of collisions (fig. 10). As the projectile initial velocity increases beyond threshold, it penetrates deeper into the target ( cf. fig. 5), dissociating more bonds and creating more fragments of decreasing size (number of atoms). This trend levels off, as expected, beyond the velocity which is required for the projectile to make its way throughout the target, i.e. about 200 \AA/ps in the present instance. For a given initial velocity above this limit, the average and maximum sizes of products decrease as the transverse dimension of the projectile gets closer to that of the target. At equality, which is the case of fig. 10, the average size of products is then smaller than 2 atoms, meaning that, as observed, there are many free C and H atoms among the fragments; also, the maximum size of the latter is about 10 atoms, meaning that the fragments are molecular rather than granular. Within the limits of validity of the present treatment, there is no reason to believe that this result depends on the initial sizes of projectile and target. We are therefore led to surmise that, in most circumstances, \emph{vaporization} is the main destruction process affecting the grain size distribution. The next section considers this issue in more detail.

\section{Grain destruction by collision}
From the previous sections, we can draw the following simplified picture. Consider a target grain of size D and a projectile of size d and initial velocity V towards the target. Let R(V) be the range of the projectile through the target material. Several generic cases must be considered for the outcome of the collision.

\begin{enumerate}
\item
 D=d
\begin{enumerate}
 \item
d$<$R: complete vaporization of both grains (cf fig. 7 and 8);

 \item
 d$>$R: part of the projectile (volume v=d$^{2}$R) penetrates into an equal volume of the target and both volumes are vaporized; formation of a crater of depth R and transverse size d in both grains;
\end{enumerate}
\item
 D$>$d
\begin{enumerate}
\item
 R$<$d$<$D: vaporized volume v= 2d$^{2}$R; cratering in both grains;

\item
 d$<$R$<$D: v=d$^{2}$(R+d); cratering in target, and projectile entirely vaporized;

\item
d$<$D$<$R: v=d$^{2}$(d+D); complete vaporization of projectile and of part of target (hole of diameter d).
\end{enumerate}
\end{enumerate}
As in fig. 8, let the \emph{destruction efficiency} of the collision, f, be defined by the ratio of the vaporized volume v to the total grain volume d$^{3}$+D$^{3}$. The different generic cases can be synthesized as in fig. 11, where V$_{c1}$ and V$_{c2}$ are critical velocities respectively defined by the conditions R=d and R=D. Assume, for simplicity, that the range R is proportional to V. Then, f has simple expressions in the three intervals of interest (see fig. 11):
\begin{itemize}
\item
10 km/s$<$V$<$V$_{c1}$: f=2Rd$^{2}$/(D$^{3}$+d$^{3}$)

\item
V$_{c1}$$<$V$<$V$_{c2}$: f=(R+d)d$^{2}$/(D$^{3}$+d$^{3}$)

\item
V$_{c2}$$<$V: f=d$^{2}$(1+d/D)/D$^{2}$(1+d$^{3}$/D$^{3}$).
\end{itemize}
\begin{figure}
\resizebox{\hsize}{!}{\includegraphics{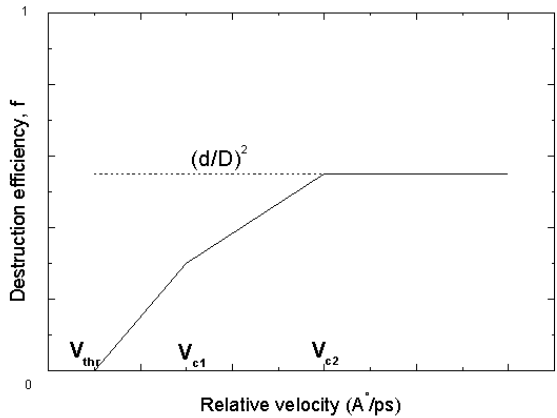}}
\caption[]{Destruction efficiency, f, of collisions between clusters of sizes D and d, as a function of initial relative velocity (cf fig. 8). V$_{\rm thr}$$\sim$10 km/s. V$_{\rm c1}$ and V$_{\rm c2}$ are critical velocities respectively defined by the conditions R=d and R=D, where R is the penetration range relevant to the model material (see text); f strongly depends on relative size of projectile.}
\end{figure}

Hence the following cursory inferences:

\begin{itemize}
\item
 No destruction occurs for V$<$10 km/s.

\item

Sticking and aggregation (coalescence) of grains occur over a small range of velocities about 10 km/s. At lower velocities there is quasi-elastic recoil. The case of a single-atom projectile stands apart because of the strong chemical reactivity of single atoms.

\item
 The collision destruction efficiency is highest for relative velocities higher than V$_{c2}$, and only reaches 1 for reactants of equal sizes.

\item
C-C and C-H bonds stand nearly equal chances of being broken.

\item
Whatever part of the grains is destroyed, is not fragmented into intermediate-size grains, but reduced to molecules of a few atoms (vaporized). This seems to question the role of collisions in producing the famous grain size distribution in r$^{-3.5}$ (see also Thebault et al. \cite{the03}). If this is confirmed, then the origin of such a distribution must be sought elsewhere.
\item
 For V=100 km/s (or 1000 \AA/ps), R$\sim$80 \AA{\ }and complete vaporization occurs only for equal grain sizes smaller than 80 \AA.

\item
Equation 2 cannot be extrapolated to higher velocities because the present treatment does not include \emph{electronic} interaction between projectile and target (see Sigmund \cite{sig69}). However, experimental and theoretical results indicate ranges increasing with projectile K.E. to powers of 0.5 to 0.75 up to hundreds of keVs  (see Schiott \cite{sch68}, Biersack and Haggmark \cite{hag80}). These results suggest that complete vaporization of the largest grains  in the ISM ($\sim0.1\mu$m) requires relative velocities of thousands of km/s.

\item
For high velocities and unequal grain sizes, part of the big grain and all of the small grain is vaporized into small molecules which carry, in the form of translational energy, most of the collisional energy that was not used for dissociation. The residual grain fragment(s) do(es) not reach temperatures high enough for it to suffer significant structural changes as compared with the initial material of the reactants. This conclusion is partly dependent on the microscopic, dynamic approach used here, as opposed to the more usual macroscopic, thermodynamic treatment. The former implicitely assumes that all translational energy carried by fly-away fragments is lost to further dissociation or heating, which is not the case in the latter. This must make a difference for the largest grains.

\end{itemize}
In conclusion, the present atomistic, non-equilibrium, dynamical approach sheds a different light on, and warrants some caution in, assessing the threshold for grain destruction and the impact of collisions on grain size distribution as well as on grain processing. This detracts nothing from the validity of the shock wave treatment in cases where most of the projectile energy is delivered in the bulk of the target by a wave travelling at a velocity of a few km/s. Such is the case when the projectile size and range in the target are much smaller than the size of the latter. Numerical simulation is also applicable in this situation, and can be very helpful in describing crater formation and temperature and pressure ditribution throughout the target as a function of time (see Cleveland and Landeman  \cite{cle92}; Aoki et al.  \cite{aok97}).

\end{document}